# Quantum wave packets in space and time and an improved criterion for classical behavior

C. L. Herzenberg

**Abstract**
An improved criterion for distinguishing conditions in which classical or quantum behavior occurs is developed by comparing classical and quantum mechanical measures of size while incorporating spatial and temporal restrictions on wave packet formation associated with limitations on spatial extent and duration.

**Introduction**

The existence of both quantum behavior and classical behavior in our universe and the nature of the transition between them have been subjects of discussion since the inception of quantum theory.

It is a familiar observation that the large extended objects that we observe in everyday life seem to engage in classical behavior, while submicroscopic objects seem to engage in quantum behavior. When a sufficient number of small quantum objects are assembled together, classical behavior seemingly invariably ensues for the combination object, and this appears to be a very robust transition. The question arises as to whether some fairly general features of our universe might account for the fact that size matters in regard to classical and quantum behavior.

Classical physics provides a deterministic description of the world in terms of localized objects undergoing well-defined motions, whereas quantum mechanics, on the other hand, provides a probabilistic description of the world in terms of waves and wave behavior. Here we will examine in a very simple manner some aspects of the circumstances in which a wave-based description of the physical world may merge into or overlap or be expressed in terms of the more familiar discrete object-based description of the physical world that we inhabit.

Since it is widely accepted that quantum mechanics is a more fundamental theory than classical mechanics, it would seem to make sense to examine semi-classical behavior from the perspective of quantum mechanics. Quantum mechanics is generally discussed in the context of a universe infinitely extended in space and time. However, we actually live in an expanding universe of finite duration in time. How might this affect the quantum waves that are present?

The description of a classical object in terms of quantum mechanics is generally approximated by the use of a wave packet composed of quantum waves. We will explore some of the general limitations on such a description introduced by a finite, limited universe.

Among the factors potentially limiting wave packet formation in a universe restricted in space and time may be the absence of very long wavelength waves, and also the absence of waves having very long periods. It would appear that these limitations might in turn limit the capability for wave packet formation in such a universe. We will examine semi-quantitatively some possible limitations on the formation of wave packets in terms of the parameters describing the extent of the universe.

The overall problem of examining possible effects of large-scale features of our expanding universe on quantum behavior and classical behavior is a dauntingly complex one that appears to be extremely difficult to address in a comprehensive manner, and it is well beyond the intent of this paper to attempt to examine this question systematically. We will introduce only very simple arguments as an aid in examining possible effects that might occur in a universe limited in time and space. We will examine temporal limitations on wave packet formation associated with the finite age of the universe and develop an improved criterion for classical behavior.

**Background: wave functions in quantum physics**

It is fairly widely accepted that quantum mechanics is a more fundamental theory than classical mechanics, so we will start from the perspective of quantum physics.

The behavior of free particles in non-relativistic quantum mechanics is described in terms of quantum wave functions that are the solutions of the Schrödinger equation that describes free particle motion.[1, 2] The basic elementary waves are the plane wave solutions each of which describes an object having a specific momentum and kinetic energy.

As is customary, a plane wave can be written as a function of space and time in the form:

$$\psi(x,t) = e^{i(kx - \omega t)} \qquad (1)$$

Here, k is the angular wave number, related to the wavelength λ by $k = 2\pi/\lambda$; and ω is the angular frequency, related to the frequency f by $\omega = 2\pi f$ and to the period τ of the wave by $\omega = 2\pi/\tau$.

Somewhat more explicitly, a plane wave solution to the free particle Schrödinger equation gives us a non-relativistic wave function describing the quantum behavior of an object having a fixed, specified momentum and kinetic energy, and can be written in the form:[2]

$$\psi_p(x,t) = e^{2\pi i(px - Kt)/h} \qquad (2)$$

Here, h is Planck's constant, p is the object's momentum, m is its mass, and $K = p^2/2m$ is the kinetic energy that characterizes the object described by the plane wave. This quantum wave function represents an object having a precise value of the momentum p and a corresponding precise value for the kinetic energy K. The object can also be described in terms of the precise spatial wavelength $\lambda = h/p$ and precise value of the frequency $f = K/h$ and of the period $\tau = h/K$.

In this connection, the quantity h/p which we have referred to as the wavelength will also be referred to as the de Broglie wavelength; while the quantity $h/m_o c$ (where $m_o$ is the rest mass and c is the speed of light) will be referred to as the Compton wavelength.

Since the probability density $P(x,t) = \psi^*(x,t)\psi(x,t)$ (where $\psi^*$ is the complex conjugate of $\psi$) is a measure of the probability for the detection of the associated object at any point (x,t) in space and time, these particular wave functions will represent objects whose probability distribution is uniform throughout all of space.

Thus, in quantum mechanics, objects having specified momentum or velocity (and corresponding kinetic energy) are described by waves of definite frequency and wavelength, and such waves describe objects whose probability as a function of location is spread out uniformly throughout space. If instead we want solutions of the free particle Schrödinger equation that can represent more localized behavior, we can construct localized wave packets using linear superposition of plane wave solutions. Thus, objects that are localized are described in quantum mechanics in terms of spatial wave packets that present a region of localized probability density. The spatial width of such a wave packet would describe the extent of localization of the object. We note in passing that, in the case of an extended object, this would provide a measure of the extent of localization or of the uncertainty in location of the center of mass of the object.

**Background: wave packets and their properties and behavior**

Wave packets can be formed by combinations of waves of specified frequency and wavelength, and can have a range of characteristics.[1, 2] In the trivial case of constructing a wave packet using only a single wave, we would as we have seen simply be describing an object whose momentum is precisely specified but whose location is not specified at all. At the opposite extreme, we could combine waves having different wavelengths so as to create what amounts to a spatial delta function, thus describing an object that is precisely localized; however, such a wave packet would need to include contributions from waves of all possible wavelengths, so that this wave packet would represent an object whose momentum would be quite indeterminate. Thus, neither extreme case would provide a completely suitable representation of a classical object.

Let's consider intermediate cases. A long wave train, limited in extent but long in compared with the wavelength, has a relatively well-defined wave number in terms of its Fourier transform, while a short wave train of limited extent would exhibit a broadened distribution of wave numbers composing its Fourier transform.[2] In wave trains of even shorter length in which the length of the wave train is comparable to or shorter than the wavelength, we are no longer dealing with an ordinary wave train but rather with a wave pulse whose principal wavelength component is comparable to the length or extent of the sampling of the wave train.

Wave packets exhibit a quite general feature that the degree of localization of a wave packet in space making use of interference effects is inversely correlated with the spread of available wave numbers. The unavoidable constraints on the spatial extent and wave number content of a localized wave packet can be expressed in a relationship that the product $\Delta x\, \Delta k$ must be comparable to or greater than a number of the order of unity, where $\Delta x$ is a measure of the spatial width of the wave packet and $\Delta k$ is a measure of its spread in wave numbers. This relationship amounts to a fundamental limitation on wave behavior, as fundamentally restrictive with respect to waves as, in other contexts, the laws of thermodynamics are, or as the limiting speed of light is in other physical systems.[2]

In quantum mechanics, these considerations are systematized and the relationship of defined uncertainties is specified in terms of the more familiar form of the Heisenberg uncertainty principle, which can be stated in terms of defined spatial and momentum uncertainties by the inequality (3 wiki uncertainty):

$$\Delta x \Delta p \geq h/4\pi \qquad (3)$$

Here $\Delta x$ and $\Delta p$ are the uncertainties in spatial location and in momentum respectively. The Heisenberg uncertainty principle makes it clear that we cannot achieve a simultaneous precise specification of location and momentum using a wave packet representation to describe a classical object.[2] However, we can form wave packets with the minimum product of uncertainties allowed by the uncertainty principle; if we do so, then we require that:

$$\Delta x \Delta p = h/4\pi \qquad (4)$$

But, since $p = h/\lambda$, we find that, in magnitude:

$$\Delta p = h\Delta\lambda/\lambda^2 \qquad (5)$$

Thus, the minimum uncertainty relation would correspond in magnitude to:

$$\Delta x \Delta \lambda = \lambda^2/4\pi \qquad (6)$$

If we further specify that the uncertainty in wavelength and the uncertainty in x (both spatial distances) should be comparable in magnitude, then we find that:

$$\Delta x \approx \Delta \lambda \approx \lambda/(4\pi)^{1/2} \qquad (7)$$

or, more roughly speaking:

$$\Delta x \approx \lambda \qquad (8)$$

Under these circumstances of wave packets exhibiting minimum uncertainties, the spatial uncertainty in the location of a wave packet will be of the order of magnitude of a wavelength.

Alternatively considered, if we want the behavior of a wave packet to resemble the behavior of a classical object, we need to minimize its spread in both space and momentum space. To get near this goal, we typically need to require that the momentum eigenfunctions participating in the wave packet should group around a central momentum value characterizing the motion of the object. If a wave packet is formed with positive momentum components with values centered around the central momentum, we can form a relatively narrow packet using a range of values comparable to the central momentum itself; if that approach is used, the wave packet width is typically of the order of a wavelength corresponding to the central momentum value.[2] Using either of these approaches or comparable means of forming a semi-classical wave packet, we find that wave packets that describe objects of reasonably well-defined velocity or momentum typically have as a principal component a wave with a wavelength that corresponds closely to the classical momentum of the object, and a width comparable to the wavelength.

We have been describing spatial wave packets, but we could in a similar fashion consider wave packets created in time, and lasting for a finite length of time, in connection with describing quantum states of finite duration.

Thus, for example, a laser pulse of finite duration (in both space and time) could be described in terms of a wave train characterized by a primary frequency; the Fourier transform of the pulse in frequency would have a width dependent on both its frequency and the pulse duration and would accordingly not be monochromatic.[2]

Such temporal wave packets would necessarily exhibit a spread or width in energy, in conjunction with the energy-time uncertainty principle that addresses the coupled uncertainties in time and energy:[3]

$$\Delta t \, \Delta E \geq h/4\pi \qquad (9)$$

**Classicality criteria in common use**

Recent work aimed at understanding the quantum-classical transition has largely addressed such important and interesting topics such as quantum decoherence and the effect of coarse-grained measurements.[4,5] However, "rule-of-thumb" criteria of earlier origin remain in use and can assist us in distinguishing conditions of largely quantum behavior from regions of largely classical behavior on a "rough-and-ready" basis.

Several criteria comparing quantum "size" measurements with classically recognizeable distances are in use to distinguish conditions of largely classical behavior from conditions of largely quantum behavior. These are based on comparison of characteristic quantum distances, such as de Broglie or Compton wavelengths, with classical distances, such as the sizes of classical objects or interparticle distances.[6,7,8]

Thus, one fairly widely used criterion tells us that as long as the physical size of an object is large compared to its de Broglie wavelength, the object can be adequately treated using classical physics.[6] Another such criterion in use tells us that if the classical size of an object is larger than its Compton wavelength, then a classical description is sufficient.[7] Another related criterion tells us that when the physical size of an object is smaller than its Compton wavelength, then a quantum description of the object is needed.[6,7]

While these criteria are of some value, they have their limitations, and hence it may be worthwhile to examine possible approaches that may provide improvements to these simple criteria for classical or quantum behavior.

**Classicality criteria for extended objects based on spatial wave packet properties**

In classical physics, objects have definite locations, and definite velocities. In quantum physics, as discussed earlier, we can approximate this behavior using wave packets exhibiting minimum quantum uncertainties.

Let us consider the case of a classical extended object which has approximately the same kinetic energy and momentum as a quantum object such as we have discussed, but which has not only a specific location but also a size. We will describe the size of this classical (or semiclassical) extended object in terms of a length L, which provides an approximate linear measure of its spatial extension.

If such an object is to behave more classically than quantum mechanically, it would in the previous terms be expected to be described by a spatially limited wave packet, so that its probability density would be localized rather than spread out. Consequently, its residual quantum wave structure would be expected to be limited in extent and to remain close to the object, rather than extending far out into space away from the object. A potential criterion for classical behavior then might require that the object's associated quantum wave structure be largely confined to and remain within the object's classical spatial extent. If the physical size of the object exceeded the width of its associated wave packet, the object would appear to behave fairly similarly to a localized classical object, whereas if the physical size of the object were much smaller than the associated wave packet, the object would appear to behave in a more quantum mechanical manner. Thus, if the object's classical behavior is to dominate its quantum behavior, we would expect that the object's wave structure would be largely confined within its own spatial extent or size, and that the size of the object would exceed the size of the wave packet. Hence, to provide for semi-classical behavior based on using a wave packet whose width is comparable to its wavelength, the size of the object would at the very least be expected to exceed the wavelength associated with the object:

$$L \geq \lambda \quad (10)$$

Eqn. (10) provides a criterion that is in fact fairly widely used for distinguishing classical behavior from quantum behavior, as noted earlier. This useful criterion states that if the physical size of an object is large compared to its de Broglie wavelength, then the behavior of that object can be adequately treated using classical physics.

Thus, in our familiar world of extended objects, when an object is much larger than its quantum wavelength we can for all practical purposes ignore the uncertainties introduced by its quantum nature. For large, human-sized objects, the quantum wavelength is very, very small, and we can, for example, safely ignore the wavelike uncertainty aspects of an automobile's position and velocity when we set out to cross a street.[9]

But, what about an object at rest? If we are dealing with a stationary object either in classical physics or in quantum physics, its momentum would be zero, and therefore its quantum wavelength given by $\lambda = h/p$ would be, strictly speaking, infinite. According to Eqn. (10), a stationary object would have to be infinitely large to meet the stated criterion for classical behavior. Here is a conundrum that appears to be telling us that no object at rest can behave classically.

**Considerations in regard to this classicality criterion: wave packet behavior in time**

How can we deal with this last result that seems to be telling us that, according to a commonly used criterion, no object at rest can behave classically, when our world of classical experience seems to a considerable extent to be based on stationary objects? In some sense, this criterion is manifestly incorrect, but what modifications are required, and why?

We have not yet looked in any detail at the temporal behavior of the semiclassical wave packets that we have been discussing. So far, we have been mainly treating the temporal character of quantum waves only in terms of elementary waves having sharply defined kinetic energy and hence expressed in terms of a single sharply defined frequency. But, as we are dealing with a wave packet in space, there will be a set of spatial wave numbers present, and consequently also a set of associated temporal frequencies contributing to the full wave packet.

Furthermore, there can also be expected to be an intrinsic spread in frequencies resulting from any limitation in time associated with the formation of the wave packet. We have been examining spatial wave packets, but, similarly, we need to consider wave packets created in time, and lasting for a finite length of time, in connection with describing quantum states of finite duration. Such wave packets would, as mentioned earlier, necessarily exhibit a spread or width in energy, in conjunction with the other form of the Heisenberg uncertainty principle that addresses the coupled uncertainties in time and energy that we noted earlier.[3]

If as is currently widely accepted, our entire universe originated with the Big Bang a finite length of time ago, then it would appear that the duration of all objects and quantum wave functions may have a upper limit determined by the lifetime of the universe. Presumably all objects and their wave functions were not present prior to the Big Bang, and developed since that time. As a result, it would appear that periodic physical processes such as those associated with wave functions of extremely low frequencies (so low that their periods exceed the duration of our universe) would not have had enough time to behave repetitively and thus would not be truly oscillatory over the duration of the universe. We will interpret the implications of this behavior as it affects, and in conjunction with, the spatial wave behavior of these objects.

To the best of our present knowledge, all objects in our universe have had a limited existence in time since the beginning of this universe in the Big Bang. Accordingly, it would seem reasonable to expect that the corresponding wave functions associated with all objects in our universe might also be considered as having been initiated at or since the Big Bang. These wave functions would presumably have been undergoing temporal oscillations for the duration of time since their creation, which would correspond to a time interval shorter than or equal to the time since the Big Bang. Thus, it would seem that all objects in our universe would necessarily be characterized at present by wave packets in time with temporal widths no longer than the lifetime of the universe.

This would have little effect on high frequency quantum waves representing objects with high kinetic energy, as these wave trains would contain extremely large numbers of oscillations. However, for objects of extraordinarily low kinetic energy, the finite duration of the universe would appear to provide a cutoff, as the wave trains describing these objects would in effect be non-oscillatory, since their frequencies would be so low that they would not have an opportunity to oscillate during the duration of the universe, so that as of the present, they would be described only by wave pulses having periods corresponding approximately to the lifetime of the universe.

Let us look at this in more detail.

In a universe that started from a specified beginning (such as the Big Bang), we will presume that quantum waves were initiated at or after the beginning of the universe. Any waves with periods exceeding the duration of the universe would have waveforms such that only a portion of a cycle would have been completed since the beginning of the universe. Hence, any such waves with periods exceeding the duration of the universe since the Big Bang would present themselves as wave pulses in time, as distinct from fully oscillatory waves. This distinction is a somewhat rough one, and therefore small numerical factors will be disregarded throughout the discussion.

We will now introduce a threshold frequency by setting the period $\tau$ (that is, the inverse of the frequency, or the time to complete a single cycle) to be equal to a value corresponding to the lifetime of the universe or the temporal interval during which the universe has been in existence, a time interval that we will designate T. Let us set:

$$\tau_{th} = T \tag{11}$$

Here, T is a measure of the lifetime of the universe. The threshold period $\tau_{th}$ specified by this equation will partition free particle quantum wave functions into those whose frequency is such that multiple periods could have occurred during the lifetime of the universe, as distinct from those with frequencies so low that they would not have completed more than a single period during the lifetime of the universe. This threshold will thereby separate the truly periodic repetitive waves having higher frequencies from lower frequency wave forms whose behavior could not have been repetitive during the temporal duration of the universe, and which would be present only as wave pulses having a dominant frequency corresponding roughly to the lifetime of the universe.

If we in effect cannot set up a fully periodic quantum wave in the form of Eqn. (1) or Eqn. (2) for these extremely low frequencies, that would seem to preclude a sharply defined quantum description describing a certain set of very low kinetic energy objects. So it would appear that such objects with extraordinarily low kinetic energies could not be represented accurately by elementary quantum mechanical waves; they would have to be described by pulse type wave packets rather than elementary monoenergetic wave functions. Objects described by these non-oscillatory wave pulses seem to be possible candidates for a special sort of behavior that is distinctively different from the simple quantum wave behavior typically ascribed to free particles in quantum mechanics.

In fact, it would appear that such objects would have to be described as wave packets in time, with a width in time corresponding to the duration of the universe. Thus it would seem that any object described by an extremely low-frequency wave would exhibit only a changing waveform rather than periodic oscillations during the time available, and thus would perhaps be describable by a roughly triangular form (or single sawtooth form) of wave pulse of width comparable to the duration of the universe. Any of these rather similar packets would be characterized by a dominant frequency corresponding to this period, the lifetime of the universe. Thus all extremely low energy objects would appear to be characterized by the same period, and thus by the same frequency and also effectively by the same kinetic energy, - this seems to be the lowest kinetic energy available in such a universe to describe a free particle in motion.

We can reexpress Eqn. (11) in terms of parameters introduced earlier to describe a quantum wave. In particular, we find for the kinetic energy value at threshold:

$$K_{th} = h/\tau_{th} = h/T \tag{12}$$

Does this proposed identification of a threshold for an effective lower bound for a kinetic energy make any sense in our world? We can use Eqn. (12) to evaluate numerically the kinetic energy characterizing an object having threshold properties in such a time-limited universe. We can use the Hubble time $T_H$ (equal to the inverse of the Hubble constant $H_o$) as a measure of the duration of our universe since the Big Bang. Using a value of 2.3 x $10^{-18}$ sec$^{-1}$ for the Hubble constant, we find for the threshold kinetic energy:

$$K_{th} = h/T_H = hH_o = 1.5 \times 10^{-51} \text{ joules} \tag{13}$$

We may note that the threshold frequency, which corresponds to a kinetic energy of about $10^{-51}$ joules, would for the case of macroscopic objects at least, seem to characterize objects that are for all practical purposes at rest. This seems to result from there being, in effect, a lowest frequency in time determined by the limitation in time which is associated with the lifetime of the universe.

But a wave packet having the threshold kinetic energy would also have an associated threshold momentum and a threshold wavelength as dominant wavelength. We can inquire what momentum and wavelength would correspond to this threshold kinetic energy; by expressing the kinetic energy in terms of momentum and wavelength we find from Eqn. (13) for the momentum and wavelength values at threshold:

$$K_{th} = p_{th}^2/2m = h^2/2m\lambda_{th}^2 \tag{14}$$

Here, $p_{th}$ and $\lambda_{th}$ designate the momentum and wavelength associated with a wave at this threshold separating temporally periodic from aperiodic wave function behavior. Accordingly, we can use Eqn. (14) to evaluate the momentum and wavelength for an object having threshold properties in such a time-limited universe.

While the kinetic energy evaluated in Eqn. (13) would appear to be a negligibly small kinetic energy, the wavelengths that correspond to it have a wide range of values depending on the masses of the associated objects. If we evaluate the wavelength that corresponds to this kinetic energy by combining Eqn. (13) with Eqn. (14), we find:

$$\lambda_{th} = (h^2/2mK_{th})^{\frac{1}{2}} = (hT_H/2m)^{1/2} = (h/2mH_o)^{\frac{1}{2}} \tag{15}$$

As an indirect consequence of the finite duration of the universe, this would in effect be the longest wavelength available to any object. This threshold wavelength (rather than an infinitely long wavelength) would appear to be a wavelength that could properly be associated with objects that are considered at rest in our universe.

We note in passing that this threshold wavelength is individualized in that it depends on the mass of each object. Furthermore, it is in all cases shorter than a length restriction that might be set simply by the size of the universe, in terms of the Hubble radius, $R_H = cT_H = c/H_o$.

Thus, we see that the limitation on the temporal behavior of wave functions associated with the finite duration of the universe seem to affect the spatial behavior of objects by in effect imposing a limitation on the size of the de Broglie wavelength associated with very low energy objects.

**An improved criterion for classical behavior**

Thus, for an object that, practically speaking, is at rest in such a temporally constrained universe, an improved criterion for classical behavior would appear to be that the size of such an object would have to exceed its associated threshold wavelength:

$$L \geq \lambda_{th} \qquad (16)$$

We note that if such an object were moving rather than at rest, its wavelength would be shorter than or equal to the threshold wavelength, so this criterion expressed in Eqn. (16) would be expected to remain a valid threshold for all extended objects, including those in motion as well as those at rest.

If we regard the relationship expressed in Eqn. (16) as identifying a critical size such that objects larger than this threshold size could behave in a more or less classical manner, while objects smaller than this threshold size may behave in a more or less quantum manner, we can calculate an approximate critical length that characterizes the size of an object having the threshold or critical size $L_{cr}$ which separates quantum from classical behavior by setting the critical length $L_{cr}$ approximately equal to the threshold wavelength $\lambda_{th}$, so that $L_{cr} \approx \lambda_{th}$. Then, combining Eqn. (15) with Eqn. (16), we find:

$$L_{cr} \approx (hT_H/2m)^{1/2} \approx (h/2mH_o)^{1/2} \qquad (17)$$

This result suggests that, at least roughly speaking, objects smaller than the threshold size specified in Eqn. (17) would tend to behave quantum mechanically, while objects larger than this threshold size would be expected necessarily to behave in a classical manner, at least in regard to their translational motion. According to Eqn. (17), massive objects will have smaller critical lengths, and therefore sufficiently massive objects will necessarily behave classically according to this analysis. For ordinary human-scale objects of typical densities, the critical length estimate would correspond to a distance of the order of 0.1 millimeter and the corresponding critical mass above which classical behavior would be anticipated, would be of the order of about a microgram.[10-12] While these values are extremely rough, they provide for an at least crude separation of the classical behavior of our macroscopic world from that of a submicroscopic world that largely exhibits predominantly quantum mechanical behavior.

**Summary and further discussion**

In summary, objects larger than the critical length just derived would generally be expected to behave in a classical manner as a result of limitations on quantum wave behavior in a time-limited universe as well as limitations based on spatial wave packet behavior.

The modified criterion for classical behavior that has been developed here is intended for application in circumstances of non-relativistic motion. It appears to provide an improvement over the earlier criterion of simply comparing an object's size with its de Broglie wavelength, notably in that use of the present criterion specifies that there will be classical behavior for large objects at rest. Furthermore, this criterion presents a more stringent limit than the use of the Compton wavelength to compare with an object's size, as the critical size here is always larger than the Compton wavelength of an object.

Several earlier papers, based on somewhat different physical arguments, have come up with roughly similar criteria for a threshold between quantum and classical behavior. Arguments based on the uncertainty principle in the presence of Hubble expansion; random motion in the context of the stochastic quantum theory; and wave packet expansion have all led to closely related simple criteria for a threshold size separating quantum behavior from classical behavior.[10-15]

The results of the present analysis correspond to a good approximation to these previously obtained results for the critical or threshold length separating quantum from classical behavior. The critical limit obtained here is also in at least very rough agreement with observations as to the sizes above which essentially all objects behave in a classical manner; while this criterion provides only extremely rough guidelines, this does correspond very approximately to a threshold above which we observe classical behavior in our world of familiar objects.

While we have derived here a threshold criterion for classical behavior that depends on the Hubble time, this is derived as a limiting value for all objects. It is implicit in the derivation that a more specific limit might be derived for an individual object that could depend on the actual age of the object. Thus, this approach suggests that the classical or quantum behavior of any individual object might exhibit a dependence on the age of the object, with more recently formed objects tending to behave more classically than comparable objects that have been in existence for a longer time. In accordance with this idea, quantum objects would appear to require time to establish their quantum behavior rather than exhibiting quantum behavior immediately after formation. Further exploration of this idea will be left to future work.

C. L. Herzenberg
carol@herzenberg.net